# Influence of spectra sewing on XCT measurement

## A. J. Arikkat (1 and 3), K. A. Janulewicz (1), C. M. Kim (2), P. Wachulak (1)

((1) Institute of Optoelectronics, Military University of Technology, 00-908 Warsaw, Poland, (2) Advanced Photonics Research Institute, Gwangju Institute of Science and Technology, 61500 Gwangju,  Korea Republic, (3) Present address: ASML the Netherlands)

*wachulak@gmail.com

The paper presents an analysis of the possible spectra manipulation and its consequence for the specific application of *XCT*. The focus was on the modification of the registered spectra dominantly by the sewing/stitching method. A model spectrum was created to analyse the possible behaviour of the spectral components when specifically arranged. The model and processing of real experimental data revealed that careful spectral sewing can be a very useful procedure and typically leads to improvement of the results obtained with the *XCT* technique. The results recommended also cautiousness in the choice of the applied modification and scale. In some cases gain or spectral enhancement of a part of the spectrum can be considered also as a sort of sewing, and improve the *XCT* results.

## 1. Introduction

Radiation spectra are exploited in many applications characterised by different specifications and processing methods. The parameters such as spectrum bandwidth, its shape, and intensity level have a constitutive character for those applications. When no extensive further processing is foreseen or the considered processes do not require any transformation to other domains, as frequently happens e.g. in the absorption spectroscopy [1,2], there is no serious problem in merging the spectra recorded in different (e.g. consecutive) measurements by tuning a source emission characteristics and keeping the same experimental arrangement. However, there are applications requiring complex data post-processing to extract the information contained in a hidden form in the registered spectrum. X-ray coherence tomography (*XCT*)) [3-6] belongs to this group and is a short-wavelength variant of the well-established optical coherence tomography (*OCT*)) [7,8,11], however, providing significantly higher axial resolution of the order of a few nanometers [4-6].

The sewing or alternately used term stitching technique of spectral components recorded in adjacent spectral ranges is one of the methods applied to improve the measurement's output quality or extend its range. Moreover, it brings some advantages to the experimental technique, especially in the spectral region of short wavelengths. Short wavelengths are experimentally very challenging and require specific apparatus. By applying sewing, one can reduce requirements towards the bandwidth of the available spectrometer. This leads to improvement in the spectral resolution by extending the available bandwidth of the registered spectrum. The possibility to manipulate the spectral content of the recorded signal offers more flexibility in the detection technique and reduces requirements towards the source spectral characteristics. All this works for widening the diversity of the samples and the same for enrichment of the *XCT* technique.

It is well known that the axial resolution limit of the *OCT/XCT* method depends on the radiation coherence level [7,9-13] and hence, on the bandwidth of the applied radiation. The problem of the preferable source characteristics was discussed comprehensively in the literature [8,11,14]. A wide bandwidth is a pre-condition for an increase in the method's ultimate resolution while the Gaussian spectrum shape was always considered as the most preferable one. The last preference is hardly available in the short-wavelength spectral range with typically complex structure due to frequent overlapping of the lines, bands, and the presence of the absorption edges. Many factors such as noise, dispersion, or numerical windowing/filtering limit the method's axial resolution but the battle for a broad spectrum is the first step toward resolution improvement. On the other hand, there are numerical procedures, such as pre-transform zero-padding, that lead to oversampling in the other domain of the Fourier transform (*FT*) pair. This does not formally increase resolution as no additional features can appear after transform execution but it delivers more sampling points and hence, it improves interpolation of the output or in other words, it increases the presentation resolution. The filters, non-transparent to visible radiation, are indispensable in many aspects of the *XCT* 's experimental practice, e.g. to protect the CCD sensors and cut off the parasitic scattered photons, especially of the visible radiation [3,6]. Filters usually strongly reduce the intensity of some part of the source's short-wavelength radiation spectrum, especially close to the absorption edges of material and this can be treated as a sort of specific filtering or windowing. Interestingly, the numerical filters, such as Kaiser-Bessel or Hamming window, are applied frequently in signal processing to reduce the number of artifacts in the post-processed spectrum. Sewing/stitching the signals obtained within the high-transmission areas of two different and compatible filters seems to be one of the natural and tempting methods to avoid reduction in resolution and even to increase it by increasing the signal-to-noise ratio (*SNR*). The latter depends dominantly on the signal power.

In the paper, we propose and discuss the spectral sewing method for the first time as a means of improvement for a specific application. The proposal is supported by an analysis of the consequences of such a novel procedure for the *XCT/OCT*. We analyse also some other possible manipulation variants of the recorded spectrum with the aim of improving the *XCT* measurement accuracy and verify its practicability. First, we explain the principle of spectra sewing/stitching based on a simple model. Then we process simple model spectra to learn behaviour of the sewed spectral components when undergoing the Fourier transform. Finally, we use the same approach to process real experimental data collected by applying a source emitting radiation in the range of extreme ultraviolet (*XUV*).

## 2. Problem formulation

The essence of the problem and a sketch of the stitching procedure are shown in Fig.1. Having a broad emission spectrum ($F(k)$ in Fig.1a) and being enforced to use a specific filter with the transmission curve ($F_1(k)$) significantly limiting the useful spectral width and introducing a low-transmission range (the effect is seen in Fig.1b), we can find a compatible filter of high transmission ($F_2(k)$) in this range and the effect of its application is shown in Fig.1c. These two filters, when jointly applied during the same measurement, would transmit a prohibitively narrow spectrum of the useful photon flux. However, one can record the spectra in two consecutive measurements and then sew together their high-level (high-intensity) parts e.g. at a spatial frequency $k_s$, where both signals are identical or at a very close level (marked by a dashed line in Fig.1d). The procedure can be, of course, extended to multiple sewed spectra, if needed.

In the *XCT* (as well as in the *OCT* in a non-interferometric or common-path arrangement) a signal reflected from a sample and normalised to a separately recorded reference signal does constitute the recorded spectrum of interest. The obtained signal is formally a spectral power density of the radiation and contains encoded information about the structure buried in the depth of the reflecting object. In other words, it is a transfer function of the sample, typically denoted as $H(k)$, in the Fourier optics. As the spectral power density in the spatial frequency domain is directly related to the autocorrelation function *ACF* in the space domain by the Wiener-Khinchin theorem (*WKT*), i.e. precisely by the inverse Fourier transform (*IFT*), we took the *ACF* as a measure to estimate and compare the consequences of spectral manipulation performed by different methods. In the common-path arrangement, the calculated *ACF* represents the desired signal equal to $|r(z, \phi)|^2$, where $r(z, \phi)$ is an effective reflectivity of the investigated sample [5,11,13]. This "distributed" reflectivity is a function of the reflector position $z$ and the incident radiation phase, i.e. the optical path length. It is expected that *ACF* will be narrow to conserve the sharp peaks identifying the hidden reflecting objects. The standard further processing, including dispersion correction, windowing, and phase retrieval will not dramatically influence the results within the discussed problem but these steps are indispensable in data processing to obtain a high-quality measurement result [5,8,11,15].

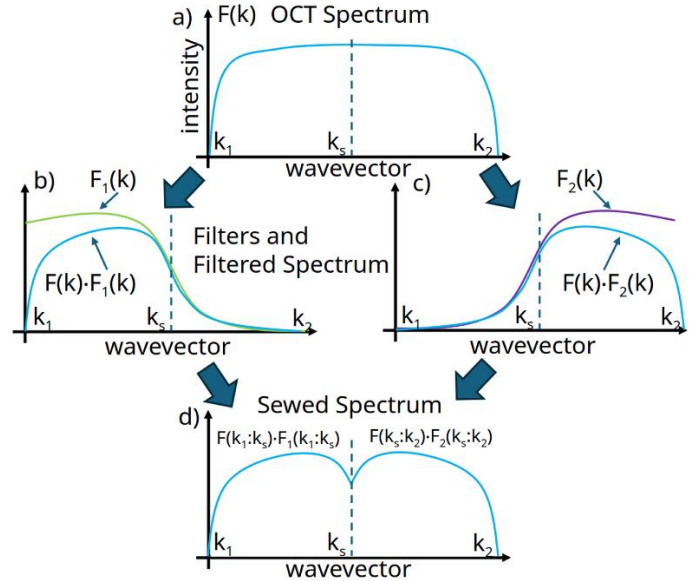

Fig. 1. Illustration of the sequence of acts leading to extension of high-level (intense) spectrum by sewing/stitching components originating from two consecutive measurements on the same sample but with compatible, as explained in the text, filters.

## 2.1. Model

We analysed a simple model based on the principle presented in Fig.1. The spectra shown in Fig.2a demonstrate the mechanisms behind the modification process (dominantly sewing/stitching) and the consequences for the *XCT* output resulting from using specific procedures. The spectra applied in the model were taken from the spectral transmissions of zirconium (Zr - red solid line in Fig.2a) and aluminium (Al - black dashed line in Fig.2a) [16] and then converted to the spatial frequency domain ($k$). Having taken the hypothetical incident photon flux as equal to one, the transmission curves (here for filter thicknesses of 200 nm and 250 nm, respectively) represented the modeled spectral irradiation level. Specific parts of the spectral curves separated by the marked $k_s$-line are identified by the corresponding description. In the model, we denoted the spatial frequency-dependent intensity as $F(k)$ with one subscript and one superscript. The subscripts $H/L$ are related to the area of high- or low-end of the spatial frequencies where the maximum of the given spectrum is localised, respectively. Thus, in our case the subscript $H$ corresponds to Zr- and that of $L$ is related to the Al-curve. The superscript $h/l$ indicates the high- or low-level signal of the given curve, respectively. The denotation principle is shown in Fig.2a. Following this explanation one can write the signals recorded within the full spectral range as

$$F_H(k) = F_H^l(k) + F_H^h(k), \quad F_L(k) = F_L^l(k) + F_L^h(k) \tag{1}$$

One can easily see that the high-level signal filtered by Al ($F_L^h(k)$) corresponds directly (it covers exactly the same spectral range) to the low-level signal obtained with the Zr filter ($F_H^l(k)$). As the spectral intervals are identical for both parts it should bring the same or equivalent spectral information about the sample structure, i.e. the reflecting layers hidden in a bulk material. For the sake of clarity, it has to be stressed that no embedded structure was considered in the model as our interest was focused on the spectrum decomposition and its consequences after the execution of the transforms. The stitched high-level parts of the signals are marked in Fig.2a by the bold lines and the resulting sewed spectrum is denoted as $F_S(k)$. The sewing point was selected at the $k_s$ value where both curves cross each other (for reference please look at the plot in Fig.1d).

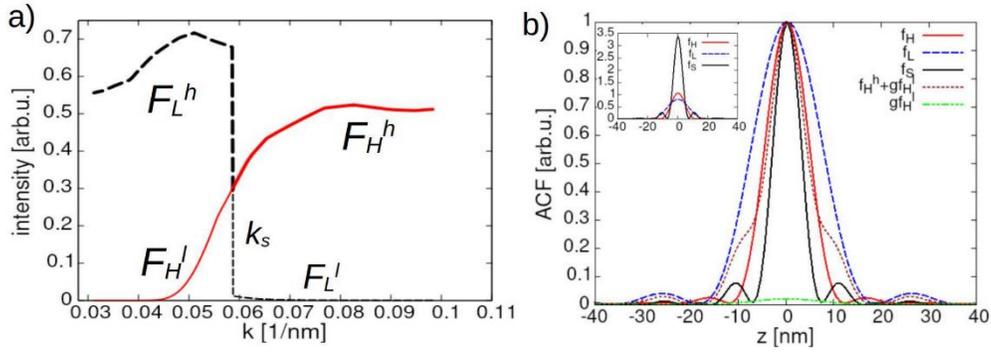

Fig. 2 a) Illustration of the spectrum modeling: two exemplary spectra converted from the transmission data of Zr and Al with the sewed high-level signals marked by the bold lines; each part of the spectrum is described according to the principle enlightened in the text; the data for spectra (Zr and Al) was taken from the *CXRO*-database [16]. b) The autocorrelation functions (spectral impulse responses) corresponding to the input $F_H(k)$ (solid red line) and $F_L(k)$ (dashed blue line) spectra are accompanied by the *ACF* obtained as a result of transformation from the spectrum sewing the high-level parts of both signals (solid black trace) and the result of transformation of the spectrum $F_H(k)$ with the part $F_H^l(k)$ enhanced by a factor of 5 (brown dotted line marked in the legend as $f_H^h+gf_H^l$). The inset shows three of these *ACF*s when not normalised.

The spectrum resulting from the sewing process can be written as $F_S(k) = F_L^h(k) + F_H^h(k)$. Application of the *WKT* to our signals resulted in the *ACF*s presented in Fig.2b with that corresponding to the sewed spectra marked by a black solid line. This plot shows that one obtains different *ACF*s for each spectrum under consideration. Generally, the *ACF*s differ only slightly in the width of their DC parts (main lobes) but the sewed spectrum is the narrowest one – an equivalent to the highest axial resolution after such an intervention into the input data [9, 10]. The ringing effect for the ACFs apart from the impulse response of the Zr-filtered ($f_H(z)$) spectrum is connected with the dynamics of the individual signal components but not with the applied zero-padding. The padding technique had to be used in our case as the original spectral bandwidth did not offer sufficient resolution in the presentation of the results.

### 2.2 Analysis

It is obvious and confirmed by many works that a broader, smoother, and high-intensity initial signal is beneficial for the *XCT*'s resolution [7–11,17]. The spectral sewing method, in the form presented above, needs more attention as even if simple and apparently correct, it uses parts of the radiation spectra filtered out by two different filters in separate recording acts. In the example presented in Fig.2a, the sewed spectrum was obtained by merging two high-level signals when the corresponding low-level parts ($F_{H/L}{}^l(k)$) were zeroed (not discarded) to keep the same spectrum total width but although with a slightly reduced effective bandwidth of each component defined as the spectral width containing nearly full (typical accepted value of 98%) of the signal power [18]. Such an operation allowed for treating all four components separately and simple mathematical operations as addition or division of the data.

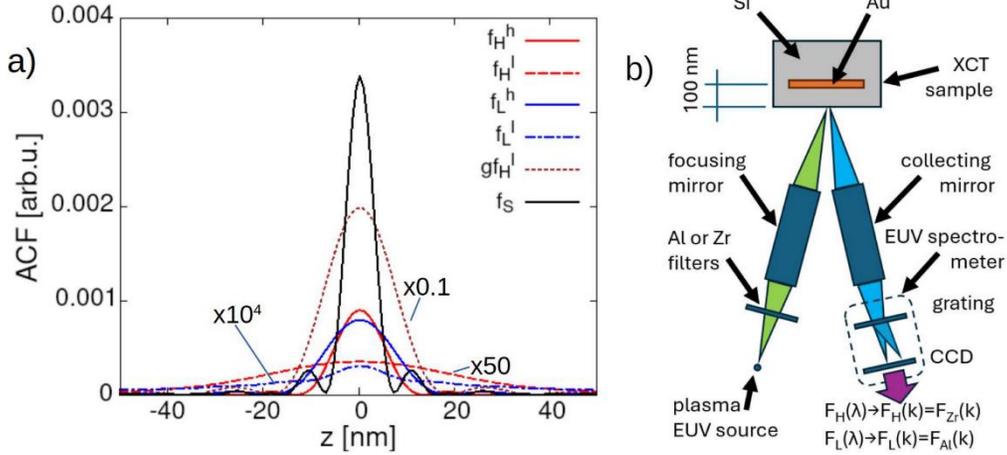

Fig. 3. The ACFs of the signal components resulting from the spectrum decomposition into low- (superscript $l$) and high-level (superscript $h$) parts as marked in Fig.2a; the impulse responses calculated from the decomposed spectra; the numbers accompanying some of the lines (e.g. ×50) indicate the multiplication factor used to present the plot within a reasonable dynamics range. b) The scheme of the experiment delivering the data used in the analysis.

Actually, the signal $F_L{}^h(k)$ replacing the low-level component $F_H{}^l(k)$ (for $k < k_s$) can be also treated as an amplified version of the latter. The amplification factor is most frequently dependent on spatial frequency $k$ and is equal to the intensity ratio of both, i.e. $G(k)=F_L{}^h(k)/F_H{}^l(k)$. Using this relation one can re-write the stitched spectrum as $F_S(k)=F_H{}^l(k)\cdot G(k) + F_H{}^h(k)$. Following the *WKT* and applying the *IFT* to the sewed spectrum one obtains the corresponding ACF of the impulse responses in the form

$$\Im^{-1}\left[F_S(k)\right] = \Im^{-1}\left[F_H^l(k)\cdot G(k)\right] + \Im^{-1}\left[F_H^h(k)\right] \quad (2)$$

The first component in the sum on the right side of Eq.(2) determines the *IFT* of a product of two frequency-dependent functions and it leads in general to a convolution of two functions in the space domain, i.e. $\Im^{-1}[F_H{}^l(k)\cdot G(k)]=f_H{}^l(z)\otimes g(z)$. Such a result can, in principle, broaden the final *ACF* in comparison to the original one on the scale dependent on both functions level. A scalar constant $G_0 = g$ is a special case of $G(k)$ and application of it can be treated as another modification procedure. This simplifies the situation as then, based on the linearity theorem of Fourier transform $\Im^{-1}[F_H{}^l(k)\cdot G_0]=g\cdot f_H{}^l(z)$, where $f_H{}^l(z)=\Im^{-1}[F_H{}^l(k)]$ is an *ACF* of the low-level signal determined in the space domain. The factor $G_0$ in the $k$-domain brings only a proportional increase in the *ACF(z)* level in the space domain, but it has serious consequences. The gain (a value of $G_0 = 5$ was used in all presented calculation results) can also be deduced from the plot in Fig.3a where the relevant *ACF* is drawn with the dotted brown line. It results in the relation $f_H{}^l(z) \ll f_H{}^h(z)$ and the weak spectral power component brings the impulse response that is noticeably broader than the part transformed from the high-level component of the spectrum. The weaker the low-level component of the spectrum (here, $F_H{}^l(k)$), the lower the level of its *ACF*. In the extreme case of $F_L{}^l(k)$ (Fig.2a) and its *IFT* equal to $f_L{}^l(z)$ (see Fig.3a), practically there is no contribution to the final result $f_L(z)= f_L{}^l(z)+ f_L{}^h(z)$. The effect depends on the enhancement factor value but it will definitely increase the difference between the interesting for us feature and the surrounding it accidental ripples. It will make the feature better visible by enhancement of the gap between the specific signature and the artefacts.

## 3. Analysis of a experimental data

The experimental data were collected during an *XCT* experiment conducted with a laser-plasma X-ray source (*LPXS*) within the scheme presented in Fig.3b. This data can be used to verify the trends extracted as the modeling conclusion. A gold film was buried nominally 100 nm below the surface of a silicon wafer. The *LPXS* emitted radiation in the spectral range between 10 nm and 32 nm. This radiation was filtered either by zirconium (200 nm thick) or by aluminium (250 nm thick) layers, and after reflection from the sample was directed by collecting optics towards an *XUV/EUV* diffraction grating creating with a *CCD* sensor the *XUV* spectrometer. More information about the source, sample, and experimental arrangement can be found in [19]. Herein, we refer only to the data necessary for the sake of the current argument. In both variants, the transfer function signals containing encoded information about the layered structure of the material were obtained by normalisation to a reference signal. The latter was registered after reflecting the source radiation from bulk silicon (Si) without any metallic layers beneath the surface. Both transfer function signals of the structure obtained in the course of this process are presented in Fig.4a.

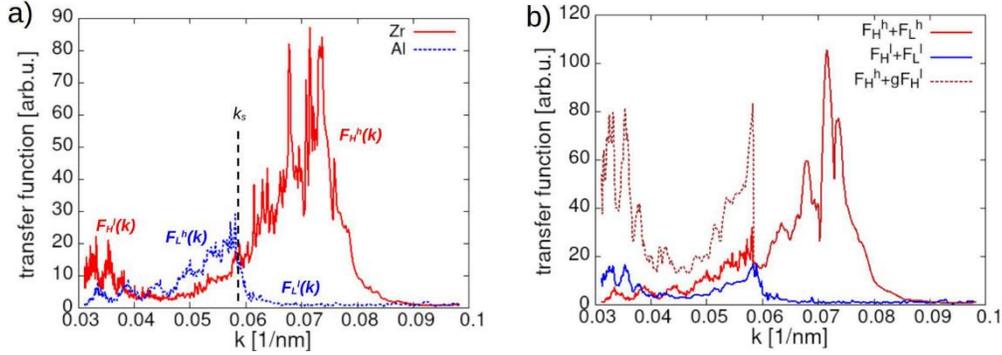

Fig. 4. a) The spectral response function of the embedded structure for both variants of filtering; b) The spectra obtained by stitching the high- (red) and low-signal (brown) parts of the spectral response functions accompanied by the spectrum obtained by replacing the low-level part $F_H^l(k)$ by its fivefold value $gF_H^l(k)$, i.e. explicitly $g=5$.

We applied the stitching technique and the new signals in both versions, i.e. merging both high-level and both low-level spectral parts, are described as $F_H^h(k) + F_L^h(k)$ (solid red line) and $F_H^l(k) + F_L^l(k)$ (solid blue line), respectively, and can be seen in Fig.4b. The sewed spectra in Fig.4b show that both combined parts are at a noticeably different intensity levels, especially those denoted as high-level ones (marked with the superscript h). This comment is valid, especially in the case of Zr-filtered radiation, where the high-level part contains also a very weak signal. As a result, the extracted *ACF*s show more complex structures. An arbitrarily chosen factor $G_0=5.0$, used in the enhancement, brought this part nearly to the level equal to the Zr-high-level part (see the brown dotted line in Fig.4b). The *IFT*s of all spectra including those described as the original input ones and those modified by stitching and enhancement of the low-level part revealed the embedded structure signals when transferred to the space domain and the result is presented in Fig.5. The collected impulse responses of the structure confirm, to some extent, the qualitative conclusions attained during modeling even if the character of the real spectra differs significantly from that of the model. All *DC* parts (the main lobes) are not surprisingly of nearly identical broadening (the *FWHM* about 10 nm) but start to differ noticeably by the beginning place of the ringing effect and the scale of it. Interestingly, the narrowest *DC*-parts of the *ACF*s show the composite spectra which came into existence either by merging high-level part of Zr-filtered signal with the enhanced low-level part of the same signal or by combining the weaker parts of both filtered signals. Characteristically, the sample impulse response (*ACF*) to the purely Zr-filtered radiation (the solid red line in Fig.5a) deflects from other *ACF*s quite early by creating a small lobe around $z=15$ nm. The further trace of this line resembles that of the lines recorded with purely Al-filtered radiation (solid blue line) and that obtained by stitching of both high-level parts of the irradiation spectra (solid black line in Fig.5a) by a small hump between 20 and 30 nm.

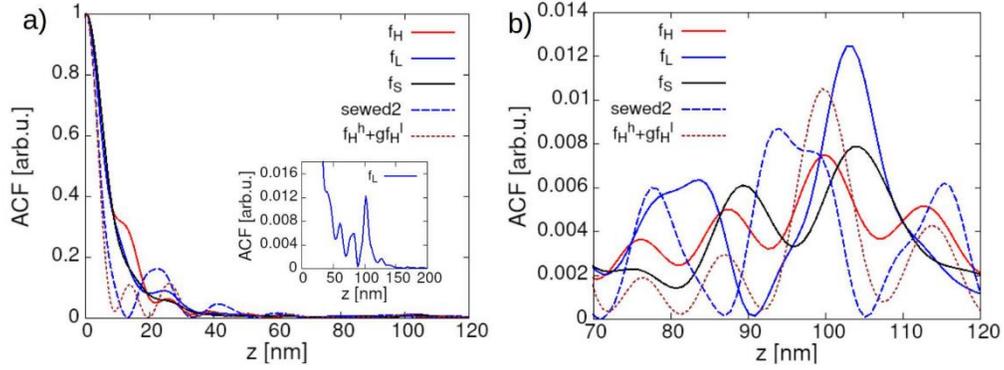

Fig. 5. The normalised impulse response functions of the embedded layered structure for different methods of spectrum modification. a) both ACF signals converted from the original spectra recorded with different filters are plotted as $f_H(z)$ (Zr filter) and $f_L(z)$ (Al filter) with the solid red and solid blue lines, respectively; the combined spectra resulted in the plots described as $f_S$ (marked with the solid black line) and "sewed2" ($f_H^l(z)+f_L^l(z)$ marked with the dashed blue line) as well as that resulting from fivefold enhancement ($g = 5$) of the weaker part of the Zr-filtered spectrum ($gf_H^l + f_H^h$, with the dotted brown line). The inset shows the trace of $f_L(z)$ with clearly visible feature, related to the embedded metallic reflecting layer. b) The normalised *ACF* signals "zoomed" in the vicinity of $z=100$ nm for the sake of a more detailed shape comparison.

The presented signals in the vicinity of 100 nm were too weak to be seen when normalised to the *ACF* maximum. The normalisation changed also the real relation between the amplitudes of the signals. Here, we concentrate on the shapes of the structural impulse responses. The inset to Fig.5a shows the impulse response $f_L(z)$ obtained by *IFT* of the full spectrum filtered by the Al-filter ($F_L(k)$). The narrow and well-pronounced peak corresponding to the position of the embedded reflector can be unambiguously identified. The most interesting from our point of view is the space area covering the $z$-axis between 70 nm and 120 nm presented in Fig.5b. It is a sort of "zooming" of this area to see more details of the impulse responses obtained by the spectra transformation. Interestingly, the original Zr-filtered signal ($f_H(z)$) gives a response of quite low quality if compared to that of the Al-filtered signal ($f_L(z)$). The main peak is accompanied by two only slightly weaker side-lobes. The origin of such a shape is not fully clear but we consider the intensity drop to a very low level for $k>0.085$ 1/nm as one of the possible sources of the effect but, definitely, it is also influenced by a relatively high level of $f_H^l(z)$, i.e. the weaker part of the transfer function. Such a low-level signal, but still of a noticeable intensity, will strengthen the ringing effect giving the observed side-lobes. In contrast, the Al-filtered signal (blue solid line plotting $f_L(z)$) a well pronounced and relatively clean peak corresponding to the searched position of the buried reflector. Characteristically, the low-level part $F_L^l(z)$ is nearly negligible (Fig.4a). This is well seen also in Fig.2a and Fig.3a. Interestingly, the signature of the buried layer is slightly shifted to the position determined with the Zr-filtered radiation. The shift is equal to $\simeq 4$ nm. The difference and influence of both spectra are visible in the best way for the stitched signals. The spectra created by sewing the high-level parts of the differently filtered signals show actually two very close peaks. One, the stronger, corresponds perfectly to the peak of $f_L(z)$ confirming very strong influence of the Al-filtered spectrum on the final character of the impulse response. The second, weaker peak, agrees with one of the side-lobes of $f_H(z)$ (Zr-filtered signal). The stitched low-intensity parts of both filtered spectra indicate the searched position in accordance with that shown by the Zr-filtered radiation. However, the resolution is rather weak and the indication is quite uncertain. The sewed spectra with the enhancement of the $F_H^j(k)$ component combined two spectral parts originating in the Zr-filtered signal and gave the buried reflector position exactly at the maximum of the impulse response of $f_H(z)$ by a clear, unambiguous peak (the dotted brown lines in Fig.5 as well as a large *SNR* due to much lower side-lobes considered as the unwanted noise. We think that the observed shift should not come from dispersion as we consider exactly the same spectral intervals and the encoded information essentially should be the same but we cannot fully exclude the complex character of the spectra under consideration and some accumulation of different phase information in two independent recording acts.

## 4. Summary and conclusions

The possibility of spectrum manipulation and its consequences for the *XCT*'s measurement resolution were analysed based on a model and data from a real experiment. The idealised model with arbitrarily selected spectra should demonstrate the modification variants, serve as an example of the denotation principle, and illustrate the consequences of the specific manipulation methods in the conjugated domain of the transform pair. This simple treatment was sufficient to obtain information about the qualitative consequences of the transfer function modification in the spatial frequency domain for the results presented as an impulse response in the space domain. Clearly, the use of filters with noticeable changes in the transmission level has a consequence similar to those resulting from numerical windowing, i.e. reduction in the method resolution by smoothing the expected sharp spikes and spectral edges. The very tempting modification method of sewing the high-level parts of the spectra obtained from two separate measurements with a compatible spectral transmission range appeared to be a useful technique but it requires a specific cautiousness. Based on the presented results, a high resolution level equivalent to the narrow impulse responses represented by the *ACF*s in the space domain requires a careful choice of well-balanced spectra to be stitched. A significant difference in the sewed spectra dynamics (intensity ratio) will introduce an unwanted, additional *ACF*'s broadening, increase in the number of sharp artefacts mimicking the true features and thus, aggravating the correct identification of the buried objects. On the other hand, big dynamics of the signals with a quite asymmetric concentration of the power carried by the signal (reduction in the effective bandwidth) can be beneficial for the unambiguous impulse response as is the case for the Al-filtered radiation but one cannot exclude problems with the resolution. The results show that simple amplification of a low-level part of the investigated spectrum brings also some benefits as it enhances the weak features but in the final effect, the right feature becomes more conspicuous.

**Funding.** The work was funded by the PRELUDIUM project (no. 2021/41/N/ST7/03198) of Poland's National Science Centre.

**Acknowledgments.** AJA is thankful to Prof. Gerhard Paulus, Drs Silvio Fuchs, and Julius Reinhard from the University of Jena, Germany for helpful discussions on the project topic as well as to Dr Piotr Nyga from the Military University of Technology and Maciej Filipiak from the Centre for Advanced Materials and Technologies (CEZAMAT) both at Warsaw, for preparation of the nano-structured samples used in the experiment.

**Disclosures.** The authors declare no conflicts of interest.